\documentclass[aps,prl,twocolumn,superscriptaddress,showpacks,longbibliography]{revtex4-2}

\usepackage{comment}
\usepackage[colorlinks, citecolor=red]{hyperref}
\usepackage{hyperref}
\usepackage[utf8]{inputenc}
\usepackage[T1]{fontenc}    %
\usepackage[english]{babel}
\usepackage[pdftex]{graphicx}
\usepackage{amsmath,amssymb,amsthm,bbm, mathtools} %
\usepackage{mathrsfs}
\usepackage[normalem]{ulem}

\begin{document}


\title{Terahertz detection within charge density wave state}


\author{Zhi Li}
\email{zhili@njust.edu.cn}
\affiliation{School of Material Science and Engineering, \mbox{Nanjing University of Science and Technology, Nanjing 210094, China}}

\begin{abstract} 
Terahertz (THz) technology enables multi-Tbps satellite communications, but conventional semiconductor detectors suffer from fundamental performance degradation above 1 THz due to the Drude limit of free electrons. Here, we theoretically demonstrate that charge density wave (CDW) materials offer a paradigm-shifting solution via their collective electronic response. We show that a static bias electric field can continuously tune the THz resonant absorption frequency of CDW states from 0 to cutoff frequency, and enhance the nonlinear rectification current by more than one order of magnitude. This unprecedented electric-field tunability makes CDW materials ideal candidates for next-generation  ultrafast THz detectors working at room-temperature. 
\end{abstract}


\maketitle

\textit{Introduction--}Terahertz (THz) frequencies (0.1–10 THz) have emerged as a transformative technology for next-generation satellite communications, offering unprecedented contiguous bandwidths that enable data rates from hundreds of Gbps to multi-Tbps ~\cite{sharma25, jiang24, song2022, CH25}. Their inherent highly directional beams further provide secure, interference-free point-to-point links and significantly improved spectral efficiency, making THz particularly well-suited for inter-satellite and satellite-to-ground communication links~\cite{shen26,Zhang26,MA2025}. For spaceborne receivers operating in the harsh radiation and thermal environments of low Earth orbit (LEO) and geosynchronous orbit (GEO), high-sensitivity THz detection is not only a performance requirement but also a critical enabler for long-distance, high-capacity satellite missions.

THz detection technologies are broadly classified into three paradigms: electronic, photonic, and hybrid electronic-photonic architectures, each with distinct trade-offs for space applications~\cite{AJAYAN,Rogalski,kutas20}. Electronic detectors, represented by Schottky barrier diodes (SBDs) and high-electron-mobility transistors (HEMTs, the dominant FET variant for THz frequencies), represent the most technologically mature approach~\cite{Lin_2020,T21,Neto05}. They are currently the workhorses of heterodyne receivers for operational satellite THz systems, offering excellent room-temperature reliability and compact form factors. However, their performance degrades significantly at frequencies above ~1 THz due to increasing parasitic losses and reduced carrier mobility~\cite{Lin_2020}, limiting their utility for next-generation high-frequency satellite links. Photonic detection approaches leverage light-matter interactions for THz sensing and offer complementary advantages to electronic technologies. Photoconductive antennas (PCAs)~\cite{Singh20}, the most widely used photonic detectors, provide ultra-broadband detection spanning multiple decades of frequency and are the gold standard for laboratory THz time-domain spectroscopy (THz-TDS) systems~\cite{Mittleman2026}. However, their historical reliance on bulky femtosecond pump lasers has hindered their deployment in space, where size, weight, and power (SWaP) constraints are extremely stringent. Another important class of photonic detectors includes quantum-well infrared photodetectors (QWIPs) and quantum cascade detectors (QCDs)~\cite{AG25,Cao04,LRB13}, which can achieve exceptional sensitivity when operated at cryogenic temperatures, with recent advances enabling near-room-temperature operation for specific narrow frequency bands. Additionally, electro-optic (EO) upconversion detectors convert THz signals into near-infrared optical signals via the Pockels effect~\cite{ZL26,EO24,TJK26}, allowing the use of mature, high-sensitivity silicon photodetectors and offering a promising path for integrated photonic satellite receivers.

Despite these advances, no single detection technology currently satisfies all the requirements for next-generation THz satellite communications. Fundamentally, this limitation stems from the Drude model of free electron gas, which governs the response of all  electronic detectors based on narrow-gap semiconductors and topological semimetals ~\cite{ZX10,Topo25,Ranjan25,XJ25,jia23,WL18}. The optical conductivity of free electrons is given by $\sigma(\omega)=\frac{ne^2\tau}{m(1-i\omega\tau)}$, which suffers severe degradation at THz frequencies, especially at elevated temperatures. As $\omega\tau$>>1 in the THz regime, free electrons become increasingly insensitive to electromagnetic radiation, imposing a fundamental physical limit on the performance of all traditional detectors at frequencies above 1 THz. Charge density wave (CDW) materials offer a paradigm-shifting solution to this long-standing challenge~\cite{chen25,xiao25,WL19}. CDW is a collective electronic ground state in low-dimensional materials~\cite{CDW23,CDW25,CDW88}, characterized by periodic modulation of electron density and accompanying lattice distortion~\cite{Loth24,KW18,CDW13}. Its intrinsic energy gap (typically 1–100 meV, exquisitely matching the terahertz photon energy range) and ultrafast collective electron response make CDW materials one of the most promising candidates for next-generation room-temperature~\cite{CDW26,KFM15,David25}, ultrafast, and high-sensitivity terahertz detectors~\cite{WNL18,Ryo21}. Unlike traditional semiconductor detectors (limited by bandgap mismatch) and thermal detectors (limited by slow response), CDW-based terahertz detectors combine the advantages of room-temperature operation, ultrafast response (ps–fs level), wide bandwidth, and low power consumption. Critically, their response originates from collective electronic excitations rather than individual free electrons, allowing them to bypass the fundamental Drude limit that plagues conventional detectors. Furthermore, the collective nature of CDW states inherently provides enhanced radiation hardness, a critical requirement for long-duration space missions. 

Despite intense recent research, two critical challenges remain unresolved for CDW-based THz detection: (1) the quantitative relationship between bias electric field and resonant absorption frequency, and (2) the mechanism for enhancing  photocurrent to achieve high signal-to-noise ratio. In this Letter, we address these challenges using a combined classical real-space and quantum momentum-space approach. We demonstrate that both the resonant frequency and nonlinear conductivity of CDW states can be efficiently tuned by a static bias electric field, providing a clear roadmap for optimizing CDW-based THz detectors.

\begin{figure}[t!]
\begin{center}
\includegraphics[width=0.48\textwidth]{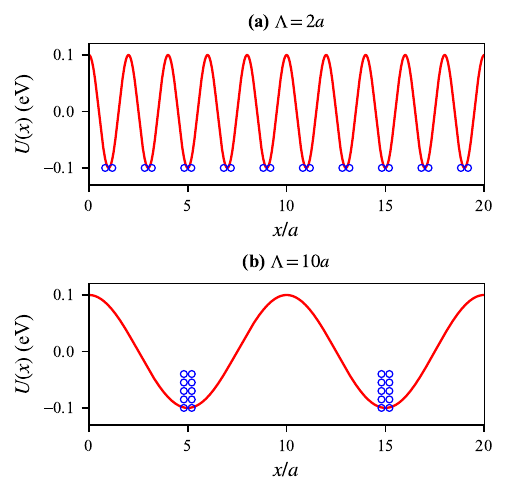} 
\end{center}
\caption{Modulation potential $U(x)=V_m\cos{(qx)}$ in crystal with (primitive) lattice constant $a$, modulation potential $V_m$=-0.1 eV, initial phase $\beta=0$, and wave vector $q=\frac{2\pi}{\Lambda}$. In normal state with vanishing modulation potential, each band is half filling. (a) In one wavelength of modulation potential $\Lambda=2a$, two electrons from neighboring lattice are bonded and trapped, and the band gap is 2$|V_m|$. (b) $\Lambda=10a$, 10 electrons develop gaped state near the bottom of modulation potential.}
\label{FIGS_cdw}
\end{figure}

\textit{Gaped state--} Van der Waals layered transition metal dichalcogenides (TMDs) are the most studied CDW materials due to their easily exfoliated atomic layers, tunable CDW phases, and compatibility with van der Waals heterostructure integration~\cite{TMD23X,TMD24L,TMDG23,TMDKis17,TMDG26}. For example, 1T-TaSe$_2$ is the most promising CDW material for terahertz detection, as it exhibits a stable commensurate CDW (CCDW) phase at room temperature (transition temperature $T_{CDW}\approx$ 473 K) with an energy gap of $\sim$60 meV~\cite{Xu_2021,TaS25,TS220,TS226,TS221,TMD24B,Li18}, perfectly matching the 0.1– 15 THz band. We start from model Hamiltonian $h_0(r)=\frac{p^2}{2m}$ descring the free electron gas in normal metallic state. Under periodic modulation potential from lattice distortion or electronic correlation~\cite{LZY21}, $U(x)=V_m\cos{(qx+\beta)}$, where $q$ is the wave vector of modulation potential, the free electron will be trapped by modulation potential, as shown in Fig. 1. With modulation potential, the single particle approximation reads $h_0(r)=\frac{p^2}{2m}+U(x)$, and total Hamiltonian reads $H=\int\psi^*(x)h(x)\psi(x)dx$. With wave function $\psi(x)=\sum_{k}a_ke^{ikx}+\sum_{k}b_k'e^{ik'x}$, the total Hamiltonian reads
\begin{align} \label{Hamitonian}
    H=\sum_k\epsilon_{a,k}a_{k}^{\dagger}a_k+\sum_k\epsilon_{b,k+q}b_{k+q}^{\dagger}b_{k+q}+   \\ \nonumber
    V_me^{i\beta}\sum_ka_{k+q}^{\dagger}b_{k}+V_{m}e^{-i\beta}\sum_kb_{k}^{\dagger}a_{k+q},
\end{align}
. After diagonalization, the energy  of lower and upper band read, $E_{d/u}=\mp d(k)=\mp \sqrt{\Delta^2+V_m^2}$, while $2\Delta=\epsilon_{a,k+q}-\epsilon_{b,k}$ (nesting fermi surfaces exist if $\Delta=0$), and the eigenvector reads $a_{k+q}=-\sin{\frac{\alpha(k)}{2}}e^{i\beta}$, $b_k=\cos{\frac{\alpha(k)}{2}}$, where $\alpha(k)$ is determined by $\cos{\alpha(k)}=\frac{\Delta}{d(k)}$ ($\sin{\alpha(k)=\frac{V_m}{d(k)}}$) and is in range of 0 (weak band dispersion) and $\pi/2$ (strong band dispersion). 
With non-vanishing order parameter $\langle \sum_ka_{k+q}^{\dagger}b_k\rangle=-\frac{1}{2}\sin{\bar{\alpha}}e^{-i\beta}$, CDW state with density modulation $n(x)=-\sin{\bar\alpha}\cos{(qx+\beta)}$ is formed. With vanishing modulation potential $V_m=0$, $\sin{\alpha}(k)$ is zero, and there is no CDW state.  The position center of occupied state $\langle x \rangle=\int_{-\Lambda}^{\Lambda}\psi^{*}(x)x\psi(x)dx$ reads
\begin{align} \label{position_center}
\langle x \rangle=\frac{1}{2\Lambda}\int_{-\Lambda}^{\Lambda}xe^{-i\beta}\sum_ka_{k+q}^{\dagger}b_k+c.c.   \nonumber
\\=-q^{-1}\sin{\beta}\sum_k\sin{\alpha(k)}
\end{align}
If the phase factor $\beta=0$, the modulation is even under spatial inversion symmetry, and the position center of wavefunction will be at the center of each unit cell, as shown in Fig. 1. Additionally, the spontaneous current is also vanishing if $\beta=0$. The detailed derivation for gaped band structure, charge density, position center and current, are include in the supplementary materials~\cite{SM}. For simplification, we assume that we start from a CDW state with band gap, initial phase $\beta=0$ (no spontaneous charge polarization or builtin electric field) and the (even) electron number is invariant in each unit cell (the single photon energy of THz field is much smaller than the modulation potential $V_m$), and we will only consider the translational motion of the CDW condensate under relatively weaker external electronic field.

\textit{Resonant frequency--} Under static electric field, the motion of electron in each unit cell trapped by modulation potential $U(x)=V_m\cos{(qx)}$ can be described by below equation,
\begin{align} \label{Static_bias}
    m\frac{d^2x}{dt^2}=-eE_b-\frac{\partial U(x)}{\partial x}=0,
\end{align}
and the center position of electron is shifted to $x=x_b=q^{-1}\arcsin{\frac{eE_b}{qV_m}}$. There exists real solutions exist for $E_b\leq E_{b,max}=\frac{qV_m}{e}=\pi\times 10^8$ V/m ($q=\pi/a$ and $V_m$=-0.1 eV), beyond which the CDW condensate depins and slides through the lattice. Under typical electric field, $E_b$=$1\times 10^5$ ~V/m, $V_m$=-0.1 eV, and vector $q=\pi/a$, the shift of center position is about $\frac{a}{1000\pi^2}$, which is minute, compared to the lattice parameter \textit{a}. To obtain relatively large shift, larger bias electric field should be applied. The detailed electric field dependent shift of position center is shown in Fig. 2a. With static bias, the modulation potential around $x_b$ point reads $U(x=x_b+u)=V_m\cos{(qx_b+qu)}=V_m\cos{(qx_b)\cos{(qu)}}-V_m\sin{(qx_b)\sin{(qu)}}$. For weak THz signal, the oscillating amplitude $u(t)$ should be small, and the modulation potential around $x_b$ reads
\begin{align} \label{Sstatic_bias}
    U(u)=V_m\cos{(qx_b)}-qV_m\sin{(qx_b)}u-  \\ \nonumber
    \frac{1}{2}q^2V_m\cos{(qx_b)}u^2  -\frac{q^3}{6}V_m\sin{{(qx_b)}}u^3,
\end{align}
Under additional THz signal with frequency $\omega$ is applied $E_{ac}(t)=|E|e^{-i(\omega -i\eta) t}+c.c.$ (infinitesimal $\eta$ for damping), the motion of electron is determined by
\begin{align} \label{Ssstatic_bias}
    m\frac{d^2u}{dt^2}=-e[E_b+E_{ac}(t)]-\frac{\partial U(u)}{\partial u},
\end{align}
and the oscillating amplitude $u(t)=u_1(t)+u_2(t)$, includes linear and second order response to strength of THz field.  For linear response, the amplitude of oscillating electron is proportional to strength of THz signal,
\begin{align} \label{S4static_bias}
   u_1(\omega)=\frac{eE_{ac}(\omega)}{m(\omega-i\eta)^2-m\omega_0^2}=\chi^{(1)}(\omega)E_{ac}(\omega).
\end{align}
. In case of resonance, the frequency $\omega=2\pi f$ of THz signal should satisfy $\omega^2=\omega^2_0$, and
\begin{align} \label{S5static_bias}
    \omega^2_0=-\frac{q^2}{m}V_m\cos{(qx_b)}=\frac{q^2|V_m|}{m}\sqrt{1-\bigg(\frac{E_b}{E_{b,max}}\bigg)^2}.
\end{align}
Without static electric bias, $\omega_0$ defines the cutoff frequency, and it is as high as 65 THz, which is much higher than the frequency of THz communication, typical 1$\sim$10 THz. In the calculation, we adopt the static mass of electron $m=m_e$.  However, static electric bias can efficiently tune this resonance frequency. Under extremely larger electric bias $E_b=\pi\times10^8$ V/m,  $x_b=a/2$, the resonance frequency $\omega_0$=0, viz. the resonance frequency of CDW state is tunable by bias electric field. The bias electric field dependent resonant frequency $f_0=\omega_0/2\pi$ is shown in Fig. 2b, and the bias field dependent resonance frequency can vary in bandwidth of 60 THz. If resonance absorption is satisfied, i.e., the oscillating amplitude of linear response $\omega^2=\omega^2_0$, $u_1(\omega)=\frac{ie\tau}{2m\omega}E_{ac}(\omega)$ and relaxation time $\tau=\eta^{-1}$.

\begin{figure}[t!]
\begin{center}
\includegraphics[width=0.48\textwidth]{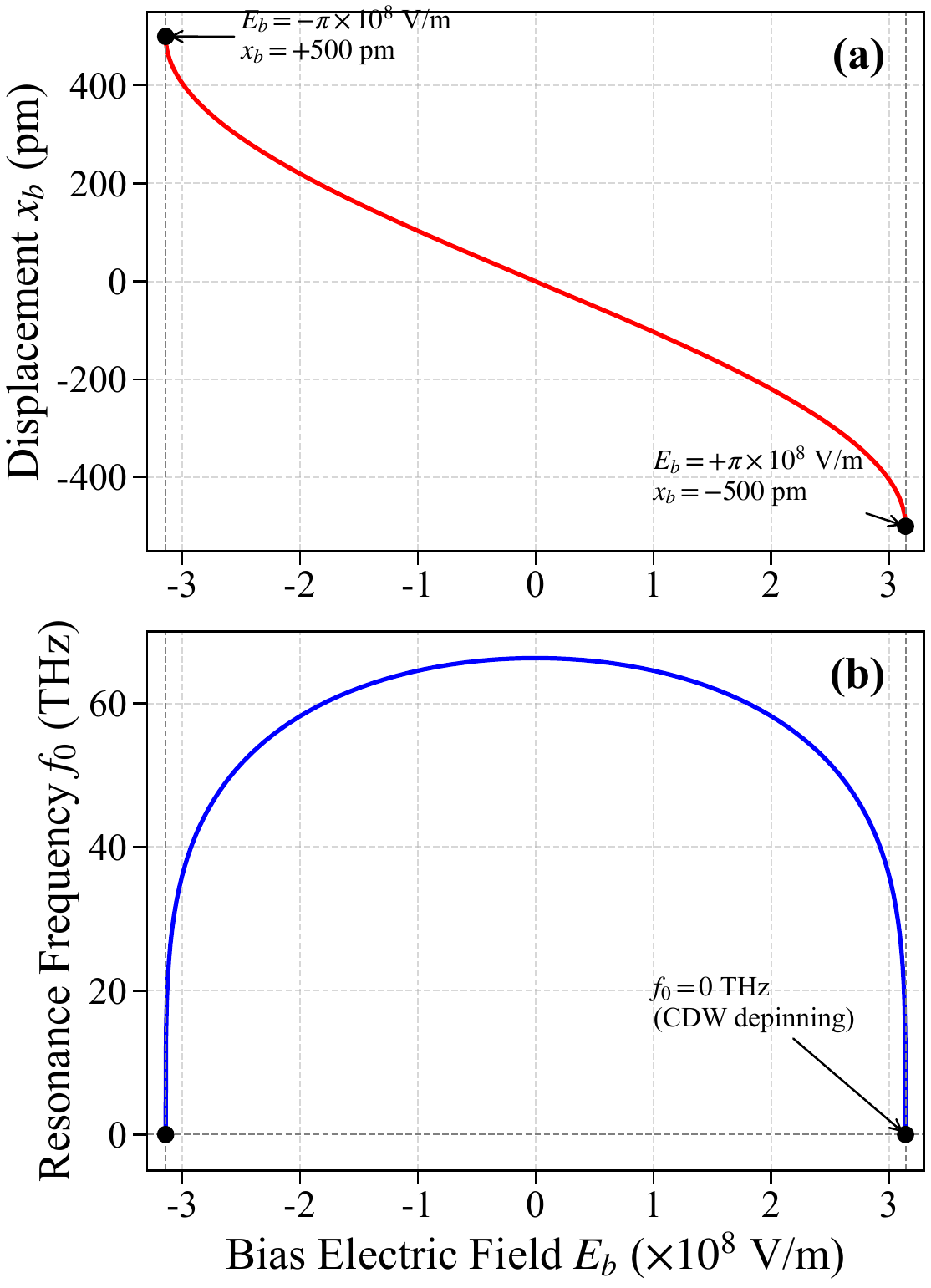} 
\end{center}
\caption{Electric field dependent shift $x_b$ of center position of electron in periodic modulation potential (a), and resonance frequency $f_0=\omega_0/2\pi$, with $q=\pi/a$, $m=m_e$ (static mass of electron), and $V_m=-0.1$ eV. }
\label{FIGS_f}
\end{figure}

\textit{Nonlinear response--} For second order response, the amplitude of oscillating electron is proportional to the square of electric field of terahertz signal, and the osculating amplitude is determined by 
\begin{align} \label{recitfing1}
  m\frac{d^2u_2}{dt^2} = q^2V_m\cos{(qx_b)}u_2 +\frac{q^3}{2}V_m\sin{(qx_b)}u_1^2(t). 
\end{align}
Since $u_1^2(t)$ term includes static component and 2$\omega$ component, $u_2(t)$ should also include two frequency components, i.e.,  $u_2(t)=u_2^0+u_2^{2\omega}(t)$, responding to shift current and second harmonic generation~\cite{SHG22,SHG89}, respectively. For second harmonic generation, the oscillating amplitude with frequency 2$\omega$ reads,
\begin{align} \label{recitfing}
  u^{2\omega}_2(2\omega)=\frac{-q^3V_m\sin{(qx_b)}u_1^2(\omega)}{m(2\omega-2i\eta)^2-m\omega_0^2},
\end{align}
and two-photons absorption happens at $2\omega=\omega_0$. For shift current current, we introduce a decay factor $e^{-\eta t}$ for terahertz field to mimic the transient current response. The shift of position center proportional to the square of THz field strength is also decaying with time as $e^{-2\eta t}$, and it reads 
\begin{align} \label{u20}
  u_2^{0}=-q\tan(qx_b)\frac{e^2\tau^2}{4m^2\omega^2}|E_{ac}(\omega)|^2e^{-2\eta t}=Ae^{-2\eta t}, 
\end{align}
when resonance absorption $\omega=\omega_0$ is satisfied. Here, $\eta$ is an infinitesimal damping parameter introduced to ensure causality, and the steady-state limit \(\eta \to 0^+\) is implied. The shift current $j=ne\frac{du^{(2,0)}}{dt}$ at resonance $\omega=\omega_0$ reads,
\begin{align} \label{shift_current1}
  j_{shift} =-\tan(qx_b)\frac{nqe^3\tau}{2m^2\omega^2}|E_{ac}(\omega)|^2e^{-2\eta t}.
\end{align}
Making use of Eq. (7), the bias dependent second order conductivity reads,
\begin{align} \label{conductivity}
  \sigma^{(2)}_{shift}=\frac{j_{shift}}{|E_{ac}|^2}=-\tan(qx_b)\frac{nqe^3\tau}{2m^2\omega^2}     \\ \nonumber 
  =\frac{eE_b}{(qV_m)^2-(eE_b)^2}\frac{ne^3\tau}{2m},
\end{align}
which is also increasing with the increase of bias field $E_b$. The calculated  static shift of position center $u^{(2,0)}$ and second order conductivity $\sigma^{(2)}_{shift}$ is shown in Fig. 3a and Fig. 3b, respectively. For zero bias electric field $E_b=0$, the position center of electron is at the center of modulation, and the system has spatial inversion symmetry. Therefore, second order response is vanishing. With nonzero bias, the position center of electron is at metastable position, and the spatial inversion symmetry is broken. With larger bias, large shift of position center will render larger second order conductivity. We conclude that the second order conductivity in the gaped state is nonlinearly dependent on the bias electric field. 

Experimentally, the CCDW phase of 1T-TaS$_2$ is characterized by a periodic lattice distortion that forms a “Star-of-David” superstructure, in which 12 tantalum atoms surround a central Ta atom in each $\sqrt{13}\times\sqrt{13}$ unit cell, accompanied by a periodic modulation of electron density. The modulation potential is also temperature and strain dependent~\cite{Gap24,WYD20}, Above 180 K, 1T-TaS$_2$ is metallic and the band gap from modulation potential should be vanishing. However, similar nonlinear dependence between bias and second order conductivity is also reproduced by the experimental photocurrent of 1T-TaS$_2$ (Fig. 2B in reference ~\cite{WNL18}), under weak bias $~E=1\times10^3$V/m. At low temperature, The photocurrent of 1T-TaS$_2$ in CDW state is studied by Sun et al, and nonlinear dependence between photocurrent and bias is observed ~\cite{WL19,WL25}. We also note that the experimentally adopted bias is usually in order of $10^5$ V/m for CDW state at lower temperature, in which much smaller than $E_{b, max}$ which is estimated with very small wave vector $q$ and relatively larger modulation potential $V_m=-0.1$eV. For 1T-TaS$_2$, the wave vector $|q|$ is seriously reduced by larger supercell and the underlying modulation potential $V_m$ is strongly temperature dependent.

\begin{figure}[t!]
\begin{center}
\includegraphics[width=0.48\textwidth]{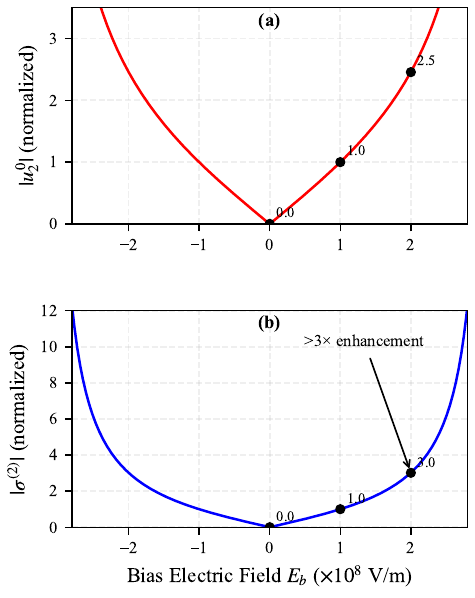} 
\end{center}
\caption{Bias-dependent nonlinear response of CDW states under resonant excitation. (a) Normalized static displacement $|u_2^0|$ as a function of bias electric field. (b) Normalized second-order optical conductivity $|\sigma^{(2)}|$ as a function of bias electric field, showing more than one order of magnitude enhancement near the bias field $E_{b,max}$. All quantities are normalized to their values at $E_b=1\times10^8$ V/m. }
\label{FIGS_sigama}
\end{figure}

\textit{Conclusions—}In summary, we have developed a unified theoretical framework for describing the linear and nonlinear response of CDW materials to THz radiation under static bias electric fields. We have shown that the resonant absorption frequency can be continuously tuned from 0 to 65 THz, and the nonlinear rectification current can be enhanced by more than one order of magnitude by adjusting the bias electric field. These results establish CDW materials as a promising platform for next-generation room-temperature, ultrafast, and tunable THz detectors.


\textit{Acknowledgements--}
This research was funded by the National Natural Science Foundation of China (Grant No. 52473304).

\bibliography{references}

\end{document}